# Aluminum nitride as a new material for chip-scale optomechanics and nonlinear optics

Chi Xiong[1], Wolfram H. P. Pernice[1,†], Xiankai Sun[1], Carsten Schuck[1], King Y. Fong[1], and Hong X. Tang[1]

[1]Department of Electrical Engineering, Yale University, 15 Prospect St, New Haven, CT 06511, USA
† Current address: Institute of Nanotechnology, Karlsruhe Institute of Technology, 76344 Eggenstein-Leopoldshafen, Germany

Email: hong.tang@yale.edu

**Abstract.** Silicon photonics has offered a versatile platform for the recent development of integrated optomechanical circuits. However, silicon is limited to wavelengths above 1.1 µm and does not allow device operation in the visible spectrum range where low noise lasers are conveniently available. The narrow band gap of silicon also makes silicon optomechanical devices susceptible to strong two-photon absorption and free carrier absorption, which often introduce strong thermal effect that limit the devices' stability and cooling performance. Further, silicon also does not provide the desired lowest order optical nonlinearity for interfacing with other active electrical components on a chip. On the other hand, aluminum nitride (AlN) is a wideband semiconductor widely used in micromechanical resonators due to its low mechanical loss and high electromechanical coupling strength. Here we report the development of AlN-on-silicon platform for low loss, wideband optical guiding, as well as its use for achieving simultaneous high optical quality and mechanical quality optomechanical devices. Exploiting AlN's inherent second order nonlinearity we further demonstrate electro-optic modulation and efficient second-harmonic generation in AlN photonic circuits. Our results suggest that low cost AlN-on-silicon photonic circuits are excellent substitutes for CMOS-compatible photonic circuits for building new functional optomechanical devices that are free from carrier effects.

## Contents

## 1. Introduction

Nanophotonic circuits made of silicon are of immense interests due to the perceived advantages of large-scale, monolithic integration with electronic circuits [1-3]. Because single crystalline substrates are available, both high-quality optical and mechanical devices [4-6] can be fabricated on a common substrate as the basis for integrated optomechanics. Due to the relatively small indirect bandgap of silicon (1.1 eV [7]), silicon waveguides are however restricted to operation wavelengths above 1100 nm. This prevents the use of shorter wavelength lasers that offer the lowest phase noise performance in the excitation of optomechanical cavities [8]. The photodetectors operating in long wavelength ranges also suffer from a relatively high dark current level. Additionally, the relative small band gap of silicon causes silicon optomechanical devices to suffer from strong two photon absorption and free carrier effects [9, 10]. Furthermore, silicon has a centrosymmetric crystal structure, which precludes important active functionalities required for on chip photon manipulation, such as nonlinear wavelength conversion and the linear electro-optic (Pockels) effect.

To complement silicon's limitations in performing active photonic functionalities, III-V semiconductors such as, InP, GaAs/AlGaAs and GaN, can be integrated on silicon chip for light generation [11] and wavelength conversion [12]. In particular, besides providing attractive optical properties, GaAs material systems allows for the fabrication of devices with high optical and mechanical quality factors. As a result, GaAs has recently found new applications in integrated optomechanical systems exploiting the combination of high optical and mechanical quality factors [13, 14]. Nevertheless, GaAs's narrow bandgap (1.4 eV) makes it susceptible to free carrier effects and limits its applications in building wideband integrated photonic and optomechanical circuits.

In this study, we present a new material system for integrated optomechanics based on aluminum nitride (AlN) thin films. Because of its large piezoelectric coefficient, AlN has been widely utilized in high-performance electromechanical devices including thin-film bulk acoustic wave resonators (FBARs), contour-mode resonators [15], and Lamb wave resonators [16]. The use of AlN as an optomechanical material allows efficient control of optomechanical systems by electrical means, and could form a basis for a new class of electro-opto-mechanical devices. Here, we introduce AlN as an attractive integrated optical material because of its excellent linear and nonlinear optical properties and mature deposition technology compatible with a variety of substrates including silicon [17-20]. Compared with GaAs, AlN provides an additional benefit by having one of the largest bandgap (6.2 eV) among all known semiconductors, and thus it not only provides suppression of two photon absorption but also allows for wide band operation from ultra-violet (UV) to infrared (IR) wavelengths. Additionally, compared with silicon, AlN has a superior thermal conductivity ($\kappa_{AlN}$=285 W/m·K) and a small thermo-optic coefficient [21] ($dn_{AlN}/dT=2.32\times10^{-5}$/K). Hence, AlN photonic devices are expected to be more tolerant to temperature fluctuations as well.

We demonstrate the versatility of the AlN platform for integrated photonics and integrated optomechanics in particular. Using CMOS (complementary metal-oxide-semiconductor) compatible nanofabrication we achieve low propagation loss in integrated waveguides and realize high-quality optical resonators at both near-infrared and visible wavelengths with a propagation loss as low as 0.6 dB/cm for the telecom C-band (between 1530 nm and 1565 nm). In low modal volume photonic crystal nanobeam cavities we measure loaded optical quality factors up to 85 000, which enables us to observe the thermomechanical motion of the nanobeam. In addition to linear photonic devices we also demonstrate efficient nonlinear optical effects in AlN by showing second harmonic generation of up to 1.7 µW of visible light on chip. Exploiting AlN's inherent Pockels effect we show GHz electro-optic modulation with electrically tunable AlN microring cavities. Furthermore, the excellent mechanical properties of the AlN thin films are demonstrated with integrated optomechanical resonators. Combining high optical and high mechanical $Q$ in a single device, we obtain GHz mechanical resonances in free-standing wheel resonators. Our platform provides a viable route towards broadband, electrically active, optomechanical devices on chip.

## 2. Substrate preparation and device fabrication

To fabricate the devices used in the current study, we first develop AlN-on-insulator substrates with a layer structure that resembles the silicon-on-insulator wafers which are widely used in silicon photonics. 100mm silicon carrier wafers are thermally oxidized to provide an optical buffer layer with low refractive index (1.45 for $SiO_2$). The buffer layer is grown to a thickness of 2 - 2.6 μm. Subsequently AlN thin films are deposited onto the oxide using a dual cathode ac (40 kHz) powered S-gun magnetron sputtering system [22]. Pure aluminum (99.999%) targets are sputtered in an argon and nitrogen gas mixture. The flow rates of nitrogen and argon gases are optimized to be around 20 and 4 sccm, respectively, to yield best stoichiometry and crystal orientation. The AlN films are deposited without external heating, thus the equilibrium wafer temperature of the substrates does not exceed a relatively low value (350°C) compared with furnace based chemical vapor deposition (normally >1000°C). As a result, the sputtering process of AlN is thermally compatible with critical CMOS processes in which no high temperature can be tolerated. The stress in the films can be tailored by adjusting the RF bias power to be around ± 75 MPa. The final AlN film is highly *c*-axis oriented with a rocking curve full width at half maximum (FWHM) of 2° at AlN's [0002] peak. Since AlN provides a comparable refractive index of 2.1 to silicon nitride's refractive index range (2.0-2.2, depending on the internal stress), photonic devices of similar dimensions can be fabricated. Therefore established design routines used for silicon nitride integrated circuits can be applied for rapid manufacture [23, 24]. Optical circuits are patterned using electron beam lithography on a Vistec EBPG 5000+ 100kV writer. Following lithography in hydrogen silsesquioxane (HSQ) resist, the exposed structures are transferred into the AlN thin film using inductively coupled plasma (ICP) reactive ion etching (RIE) in $Cl_2/BCl_3/Ar$ chemistry [12]. In order to prevent oxidization of the AlN film during the RIE process the addition of $BCl_3$ is necessary to maintain high etch rates. By adjusting the ICP and RIE parameters (i.e. the power values for both the ICP and RIE units to 350 W and 70 W, respectively, at a pressure of 5 mTorr) we preserve a target etch rate on the order of 200 nm/min, which leads to acceptable surface roughness and sufficient etch selectivity against the HSQ resist (>2:1). In order to characterize the roughness of fabricated devices after etching we perform atomic force microscope (AFM) measurements. We scan the AFM cantilever both across the waveguide surface and along the waveguide sidewall. Due to the radius of the AFM tip, the results show a rounded waveguide profile; however the sidewall roughness is still reproduced. Even though the resolution on the sidewall roughness is lower than the measurement on the top of waveguide, the accuracy is acceptable to provide an estimate of the etching quality. From the obtained data we determine an average top surface roughness of 1.2 nm rms (due to the sputtering process employed) and a sidewall roughness of 3.5 nm, which is well below the target optical wavelengths used in the current study.

## 3. Linear nanophotonic circuits

### 3.1 Determination of the waveguide propagation loss

Using the fabrication procedure established above we first realize linear photonic components in AlN. In order to assess the propagation loss of waveguiding structures we fabricate winding optical waveguides of different total lengths to extract the propagation loss. In order to preserve real estate on chip the waveguides are arranged in a long spiral pattern. An optical micrograph of a typical fabricated device is shown in figure 1(a). By adjusting the number of turns in the spiral the waveguide length can be continuously varied.

In order to eliminate loss effects due to the grating couplers we design the device for balanced measurements with an additional control port as shown in figure 1(a). By using a three-port device the transmission through the reference arm and the long spiral waveguide can be recorded simultaneously. Subsequently the coupler influence is removed by normalizing the transmission spectra recorded in the two arms.

In order to characterize the waveguide properties we fabricated devices with waveguide lengths from 300 µm to 2 cm. Measuring the attenuation through the long waveguide allows us to extract the waveguide propagation loss due to residual surface roughness. We normalize the light coupled out of the right output coupler to the light extracted from the central calibration port in order to remove loss contribution due to the input coupler. The results for a typical device are shown in figure 1(b), where we plot the obtained propagation loss at a fixed wavelength in dependence of device length, as measured from a set of different devices. A linear fit to the experimental data reveals the propagation loss per unit cm (red solid line).

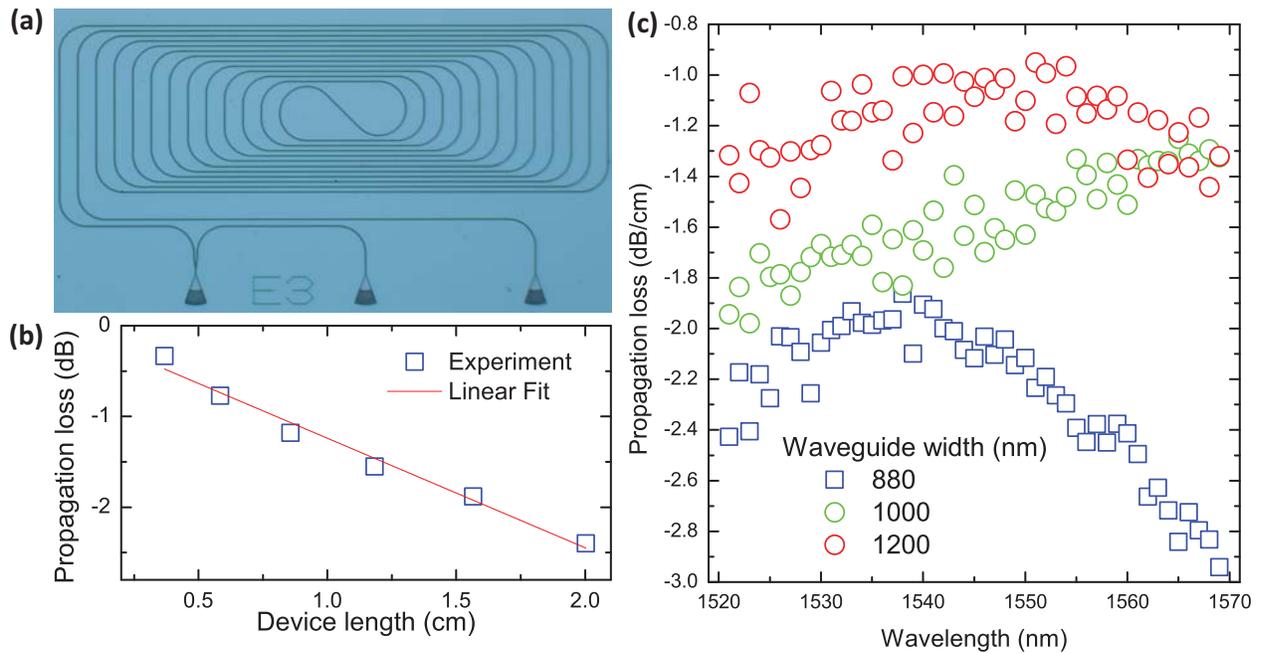

**Figure 1**. (a) An optical micrograph of the device used to characterize the waveguide propagation loss. In order to realize long structures, the waveguides are laid out in meander form. A separate calibration port is employed to account for fiber-to-chip coupling loss. Light coming from the input port (left most coupler) is split using an adiabatic 50/50 splitter. One path is led directly towards a separate output port (center coupler) for calibration purposes. The second path is arranged in a long spiral pattern in order to save chip area and reduce the device footprint. The bend radius is kept at 40 µm in order to avoid loss due to bending scattering. (b) The measured loss value in dependence of device length for a typical device. A linear fit to the data (red line) yields the loss per device length. (c) The propagation loss measured in dependence of wavelength. Three sets of data are obtained for waveguides with a nominal width of 880 nm, 1000 nm and 1200 nm.

Because the waveguide dispersion affects the propagation properties of the waveguide we measure the propagation loss in dependence of wavelength. Results for several devices are collected in figure 1(c). Each data point is obtained from a measurement result as shown in figure 1(b) as the fit to the slope of the loss curve. We measured waveguides with three different nominal widths from 880 nm to 1200 nm. The lowest propagation loss is obtained for the widest waveguide at wavelengths around 1550 nm. In this case we obtain propagation loss as low as 0.8 dB/cm. For the thin waveguides, the best loss value is measured to be 1.9 dB/cm around 1540 nm. Compared to silicon photonic waveguides the low loss values illustrate, that efficient waveguiding can be achieved on our aluminum nitride photonic platform.

*3.2 AlN optical microring resonators*

In order to confirm the loss results measured with the spiral waveguides we employ on-chip resonators for an independent loss measurement. Low propagation loss allows for the realization of high optical quality factors in suitable geometries. Therefore we analyze two design strategies: microring resonators and photonic crystal cavities. Both types of cavities have been widely used in cavity optomechanical systems made from silicon [5, 25] and silicon nitride [26, 27]. When coupled to nanophotonic waveguides, these designs can be conveniently analyzed in integrated photonic circuits.

We characterize the performance of microring resonators through transmission measurements. In order to measure the transmission past the cavity, focusing grating couplers are used to couple light into and out of the chip [28]. Two sets of couplers are optimized for IR and visible operation by adjusting the period of the grating. The couplers focus light into input waveguides, which allow for transmission measurements at both IR (~1550 nm) and red light (~770 nm). The resonators are evanescently coupled to feeding waveguides to allow for the optical characterization. Light from two different tunable laser sources New Focus 6428 (NIR) and 6712 (Red) is guided to an optical fibre array that comprises two pairs of single mode fibres for 1550 nm and 770 nm, respectively. Incoming light is adjusted with polarization controllers and collected again at the output port where it is detected by a low-noise InGaAs photoreceiver (New Focus 2011) for the 1550 nm light and a Si photodetector (New Focus 2107) for the 770 nm light.

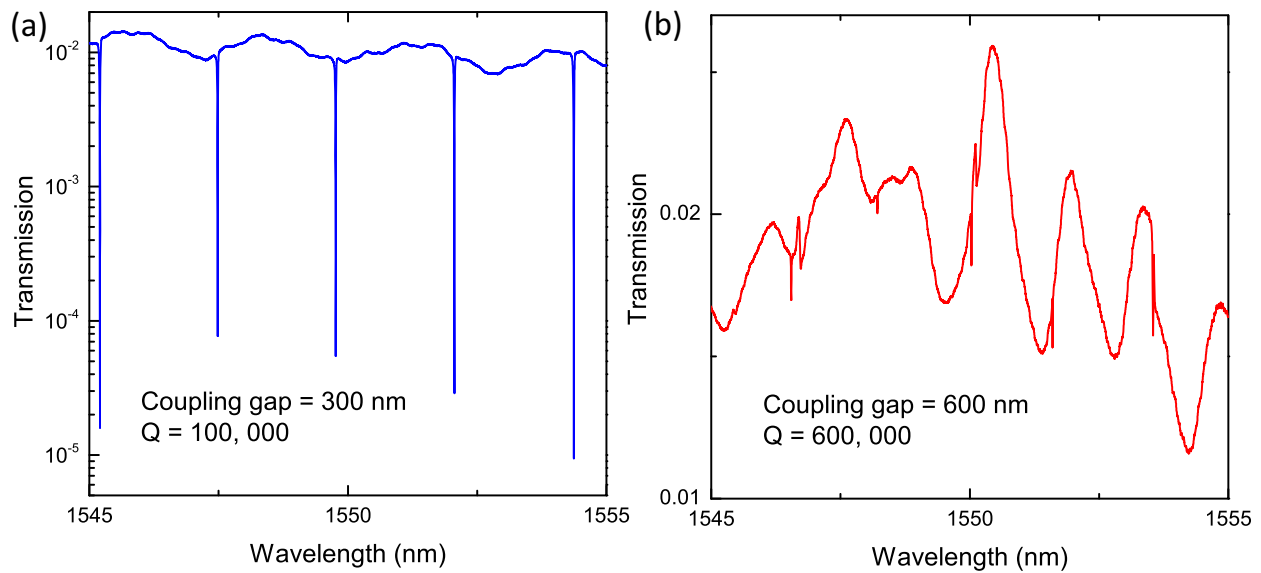

**Figure 2**. (a) The optical transmission spectrum of a critically coupled ring resonator with high extinction ratio of 30 dB for 1550 nm light input. In this case the coupling gap is 300 nm and the measured optical $Q$ is on the order of 100 000. (b) The optical transmission spectrum of a weakly coupled ring resonator with coupling gap of 600 nm and fitted optical $Q$ of 600 000.

We designed and fabricated several sets of optical ring resonators working around 1550 nm and 770 nm. A series of ring resonators with a diameter of 200 μm were used to characterize the optical quality factor at telecom wavelengths (1530 nm-1565 nm). The thickness of the AlN films is fixed at 330 nm. In order to study the coupling behavior of the resonators the coupling gap is varied from 200 nm to 1000 nm, allowing the examination of ring resonators in both the overcoupled and weakly coupled regimes [29]. The transmission measurement of the device at 1550 nm input shows optical resonances with a free spectral range of 1.9 nm, corresponding to a group index of 2.01. When the coupling gap is optimized for critical coupling (300 nm) we find high extinction ratio of up to 30 dB as shown in figure 2(a). The transmission spectrum is enveloped by the profile of the grating coupler, which provides 3dB bandwidth of roughly 50 nm. The small oscillations are due to the Fabry-Perot cavity which is formed between the grating couplers as a result of residual back reflection. When the coupling gap is further increased to 600 nm the ring is operated in a weakly coupled regime, leading to improved optical $Q$-values which are closer to the resonator's intrinsic quality factor at the expense of reduced extinction ratio. For weakly coupled devices we measure a minimum cavity linewidth of 2.6 pm, corresponding to high optical $Q$ of 600 000 (figure 2(b)). From the fit to the resonance dip the propagation loss of the nanophotonic waveguides can be estimated [30]. Using the expression $\alpha = 10 \log_{10} e \cdot 2\pi n_g / Q_{\text{int}} \lambda$ (where $Q_{\text{int}}$ is the intrinsic quality factor, $\lambda$ the wavelength and $n_g$ is the group index), we determine a propagation loss of $\alpha = 0.6$ dB/cm, which is on par with state-of-the-art silicon nitride photonic devices. The extracted propagation loss is also in good agreement with the values determined by measuring the attenuation using nanophotonic waveguides with different lengths as described in the previous section.

We next perform transmission measurements on devices designed for a wavelength range around 770 nm. In order to study the coupling dynamics directly through visible inspection we use the device geometry shown in figure 3(a). Here, we design our visible ring resonator to have a radius of 20 μm and a waveguide width of 1 μm to allow for direct comparison with the NIR results. The device comprises two set of input waveguides, which allow characterization both in the visible and the near-IR wavelength range. The 770 nm feeding waveguide, kept at a narrow width of 350 nm, is wrapped around the ring resonator over a length of 20 µm to facilitate efficient coupling into the ring resonator. The NIR feeding waveguide is kept at a width of 1 µm and a coupling gap of 450 nm, so that the ring is operated in the weakly coupled regime in order not to disturb the resonator at visible wavelengths. At NIR wavelengths we measure optical quality factors of roughly 250 000, which are slightly lower than the previous devices due to the reduced ring radius.

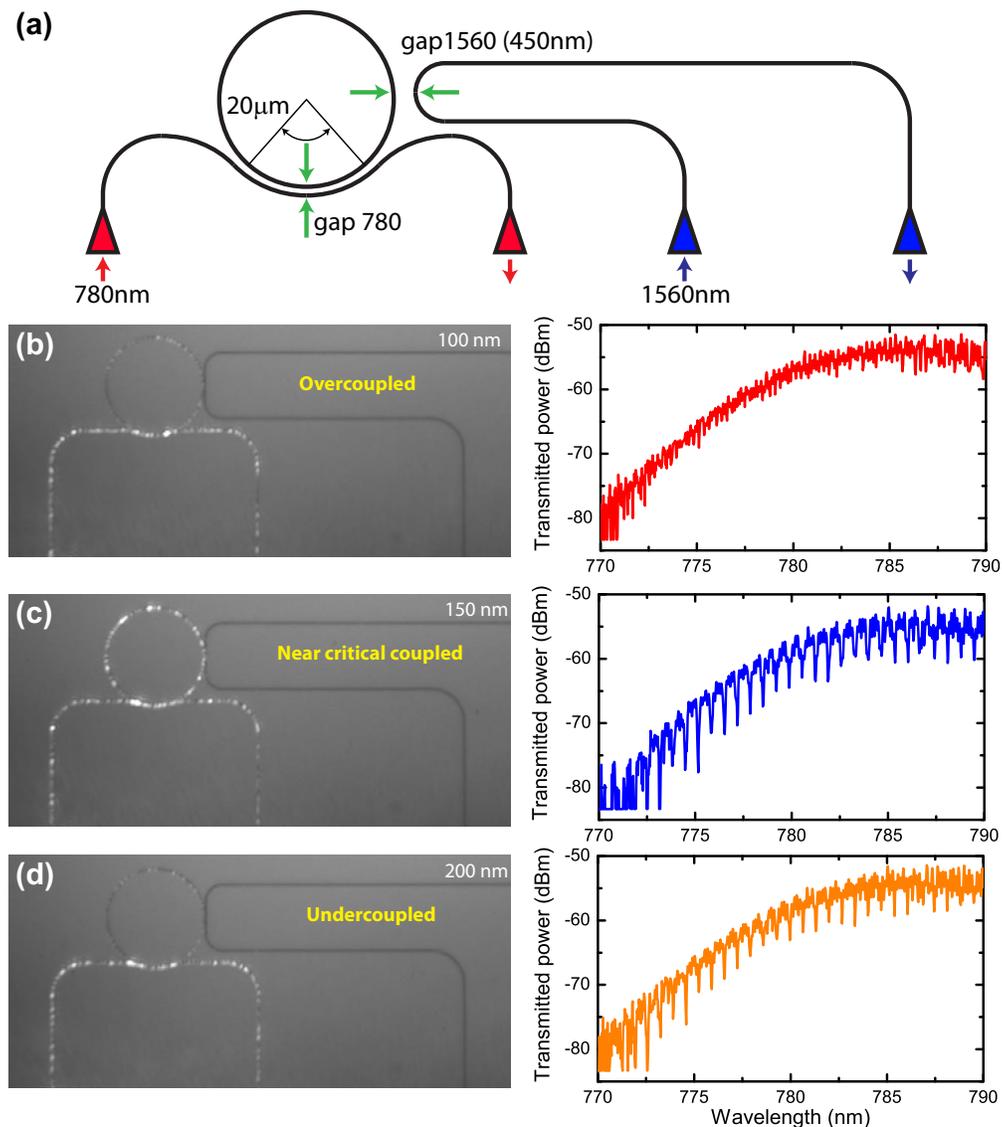

**Figure 3**. (a) The device layout used to characterize the optical response of the ring resonators in the visible wavelength range. The pulley waveguide is wrapped around the ring resonator over a length of 20 µm. (b) A device operated in the overcoupled regime. In this case the coupling gap

is 100 nm. On an optical micrograph the waveguide and ring glow. In the transmission spectrum hardly any resonances are observable due to the low extinction ratio. (c) A device with a coupling gap of 150 nm, in the near critical coupled case. Optical resonances with an extinction ratio of up to 10 dB are observed. (d) The undercoupled case, where the optical resonances sharpen, but the intensity in the ring is reduced with lower extinction ratio.

When coupling visible light into the ring resonator, the light scattered out of plane at the residual surface roughness can be collected with a CCD camera. We show both CCD images and spectral measurements side by side in figures 3 (b)-(d). When the visible feeding waveguide is close to the ring resonator, the device is operated in the overcoupled regime. In this case, the optical $Q$ is low, which implies that light travelling inside the ring resonator escapes quickly from the ring. Hence the intensity inside the ring resonator is low as shown in figure 3(b). In consequence the extinction ratio is also low, as evident from the spectral scan shown. When the coupling gap is increased further the critical coupling regime can be reached. In this case light coupled into the ring is able to circulate for several round trips leading to enhanced intensity. Accordingly the extinction ratio in the spectrum increases as shown in figure 3(c). Upon further enlargement of the coupling gap the device is operated in the weakly coupled regime, which implies that only little optical power is coupled into the ring. In this case the scattered intensity from the ring is also weak and the extinction ratio of the resonance in the spectrum reduced, while the optical $Q$ is increased. In the undercoupled regime both the linewidth of the resonances and the extinction ratio are reduced. A quality factor of 30 000 (obtained from a fitted linewidth of 26 pm), which is expected to be closer to its intrinsic value, leads to estimated propagation loss at visible wavelengths of 20 dB/cm. The increase in propagation loss is expected, given that the residual surface roughness provides more significant Rayleigh scattering ($\sim 1/\lambda^4$) at shorter wavelengths. Furthermore, both the refractive index and the material absorption of AlN increase with decreasing wavelength. As a result, the optical mode experiences a stronger refractive index contrast and is thus more susceptible to roughness scattering.

From the visible inspection the coupling behavior of the fabricated devices can be directly studied and compared to more indirect measurements based on transmission characterization. The combination of both allows us therefore the specifically engineer the coupling dynamics of our AlN resonators by adjusting the coupling gap.

*3.3 1D AlN photonic crystal nanobeam cavity*

While microring resonators provide high optical quality factors, their performance in optomechanical devices is often limited due to the large modal volume spread across the circumference of the ring. Photonic crystal (PhC) cavities on the other hand allow for shrinking the modal volume $V$ to near the fundamental limit of $V = (\lambda/2\mathrm{n})^3$. In particular one-dimensional (1D) PhC cavities offer exceptional quality factor to modal volume ratios ($Q/V$), relative ease of design and fabrication, and show much potential for the exploitation of optomechanical effects [5, 27]. Here we realize high $Q$ 1D photonic crystal cavities in free-standing AlN nanobeams by employing tapered Bragg mirrors as shown schematically in figure 4(a). The cavity is integrated into an on-chip photonic circuit and coupled to a feeding waveguide for convenient optical access as described above for the ring resonators. In our layout the cavity region is enclosed between tapered PhC Bragg mirrors, in which the primary lattice constant of the PhC is tapered down parabolically towards a secondary, smaller lattice constant in the cavity region [31]. The cavity is patterned into a free-standing AlN nanobeam with a height of 330 nm. Using a finite-difference time-domain (FDTD) method, we arrive at an optimized design with a primary lattice constant of $l_l$=560 nm which is tapered down to the cavity region parabolically over a length of 14 lattice periods to a value of $l_2$=430 nm (see figure 4 (a)). The ratio of the hole radius to the PhC lattice constant is kept at *f*=*r*/*l*=0.29, while the width of the nanobeam is set to 975 nm. By varying the distance between the inner-most holes, intrinsic optical quality factors up to $1.1 \times 10^6$ are found numerically. Figure 4 (b) shows the simulated optical mode profile for the fundamental cavity mode in the *x*-component of the electric field.

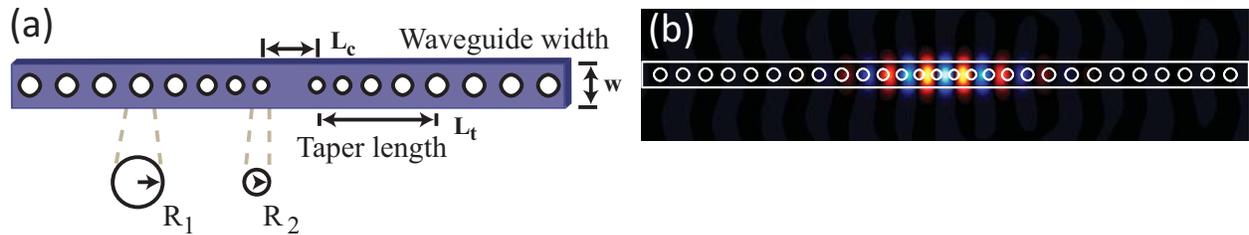

**Figure 4**. (a) The design of the one-dimensional AlN photonic crystal cavity. The hole radius is tapered parabolically from $R_1$ to $R_2$. (b) The simulated optical mode profile for the fundamental cavity mode. Shown is the *x*-component of the electric field.

The optical properties of the fabricated device (as shown in the optical and SEM images in figures 5(a)-(c)) are investigated by measuring the transmission through the feeding waveguide. Because the optical mode profile of the nanobeam is highly concentrated in the cavity region, the feeding waveguide is laid out in an arc design to only couple to the nanobeam near the point of highest intensity. By measuring the transmission profile of the feeding waveguide the cavity properties can be extracted. A clear dip in the transmission spectrum indicates the cavity resonance wavelength as shown in figure 5(d). By varying the distance between the feeding waveguide and the cavity we can operate the cavity in the overcoupled up to undercoupled regimes to adjust the extinction ratio and *Q* factors. In the near critical coupled regime we find extinction ratio of 15 dB and *Q* factor of 85 000.

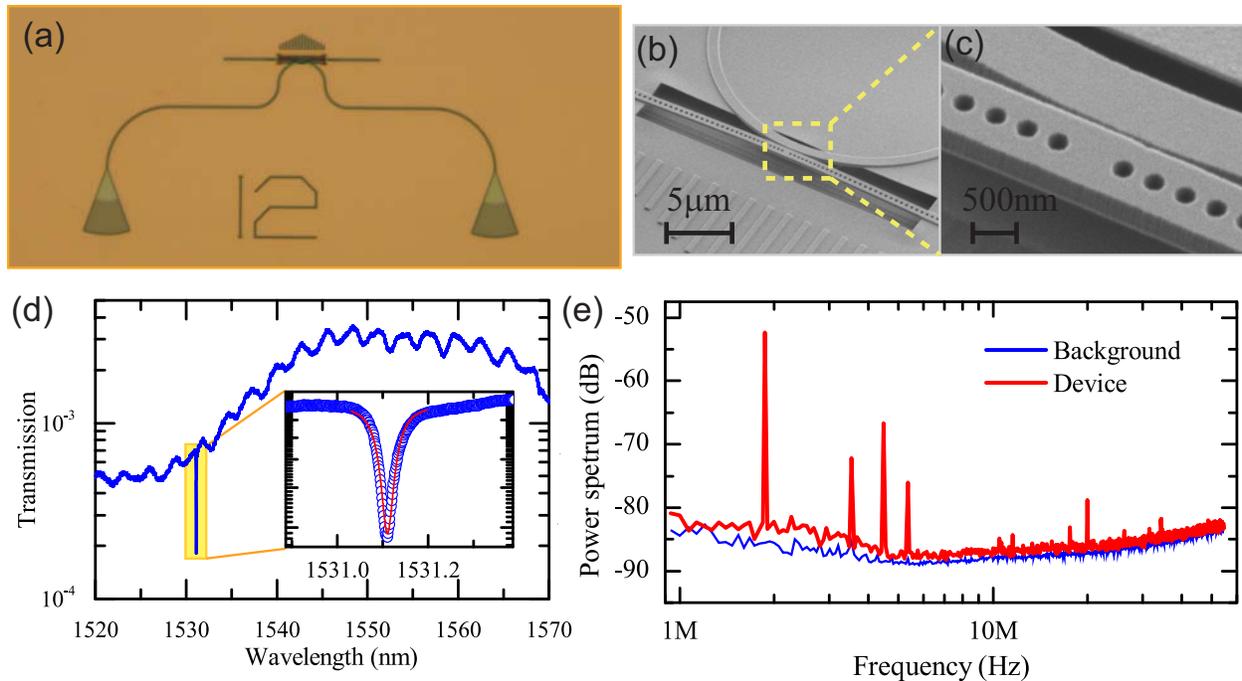

**Figure 5**. (a) An optical micrograph of a fabricated photonic circuit with grating coupler input ports, nanophotonic waveguide, and the released cavity region. (b) A SEM picture of the released waveguide region, showing the PhC nanobeam as well as the input waveguide. (c) A magnified view of the cavity section of the waveguides. (d) The measured transmission spectrum of a fabricated device, showing the cavity resonance at 1530 nm. The Lorentzian fit to the data reveals an optical $Q$ of 85,000. (e) Thermal noise spectrum of the nanobeam. The red line is the response from the device when cavity is slightly detuned, and the blue line is the spectral analyzer background.

The sharp linewidth of the cavity optical resonance allows us to monitor the thermomechanical motion of the nanobeam. We tune the input laser wavelength to the slope of the cavity resonance and monitor the transmitted optical signal with a fast photodetector. As shown in figure 5(e) we can readily identify the mechanical modes of the vibrating nanobeam up to frequencies of 50 MHz. The mechanical resonance peaks rise above the previously measured background spectrum (blue curve), which is measured when the laser wavelength is moved away from the cavity resonance wavelength. The mechanical motion modulates the distance between the feeding waveguide and the cavity, thus changing the cavity resonance condition and therefore the transmission past the device at fixed wavelength. Because of the small modal volume of the cavity and the high optical quality factor, the readout scheme is sensitive to low-power monitoring.

## 4. Nonlinear AlN photonic circuits

While AlN is a promising material for linear optical applications, it also offers nonlinear optical functionality. On-chip nonlinear optical effects have important applications in controlling and routing photons on chip [32, 33]. In particular, for cavity optomechanics, which studies the linear and nonlinear interaction between light and mechanical motion, recent progress has advanced to the level that single photon effects become relevant [34, 35]. This demands complementary on-chip techniques for manipulating photons at high speed of mechanical vibration and wavelength conversion that allows coherent three-wave interactions between photons and phonons. Therefore combining optomechanics with nonlinear optics could be a future direction for studying optomechanics in the quantum regime, either at single photon levels [36] or using continuous variables [37]. Here, in addition to linear nanophotonic devices we also exploit the inherent second order optical nonlinearity of AlN. We investigate the $\chi^{(2)}$ nonlinearity for second harmonic generation, as well as for electro-optic modulation. Due to the broadband transparency of AlN, both approaches can be used over a wide wavelength range.

*4.1 Second order optical nonlinearity in AlN waveguides*

Because AlN offers high second order nonlinearity [38], efficient wavelength conversion can be achieved. Here we aim at demonstrating second harmonic generation in which the wavelength of photons is doubled when the phase matching condition is fulfilled using the photonic circuit shown in figure 6(a). The ridge waveguides are designed to support only the lowest order of transverse electric (TE) and transverse magnetic (TM) modes at telecom wavelengths. In the case of nanophotonic waveguides, the phase matching condition implies that the wave-vectors of the pumping optical mode and the second harmonic (SH) mode at twice the frequency must obey $k_{SH} = 2k_P$, meaning that the effective indices of the pump $n_p$ and the SH mode $n_{SH}$ must be equal [39].

 In order to utilize the largest element in AlN's second order susceptibility matrix ($d_{33}$), pump light with a polarization aligned to the *c*-axis of the AlN crystal (TM-like) is required for the highest conversion efficiency. This out-of-plane crystal direction is aligned during the AlN growth, while the in-plane crystal orientation is random and thus largely isotropic. Therefore we optimize the polarization of the pump light before entering the optical circuit using polarization controllers. Then as illustrated in figure 6(a), amplified pump light from a CW telecom tunable diode laser source is launched into the waveguides from the left to excite a TM-like guided optical mode. A power monitor port for the pump light (center) and a detection port for 775 nm light (right coupler) are split out near the output end of the waveguides.

 Figure 6(b) illustrates numerically calculated effective indices of the fourth TM mode at the SH wavelength and the fundamental TM mode at 1550 nm versus the waveguide width. As the plot shows, for fundamental light at 1550 nm (red line) we have a phase matching point near a waveguide width of 1.06 µm, where the calculated effective index is equal to that of the fourth order mode at 775 nm (blue line). The corresponding modal profiles for the fundamental (bottom) and higher order modes (top) are shown in the insets of figure 6(b). We observe experimentally that only for a small range of waveguide widths, the SH can be generated efficiently when the phase matching condition is fulfilled [40], confirming the theoretical predictions.

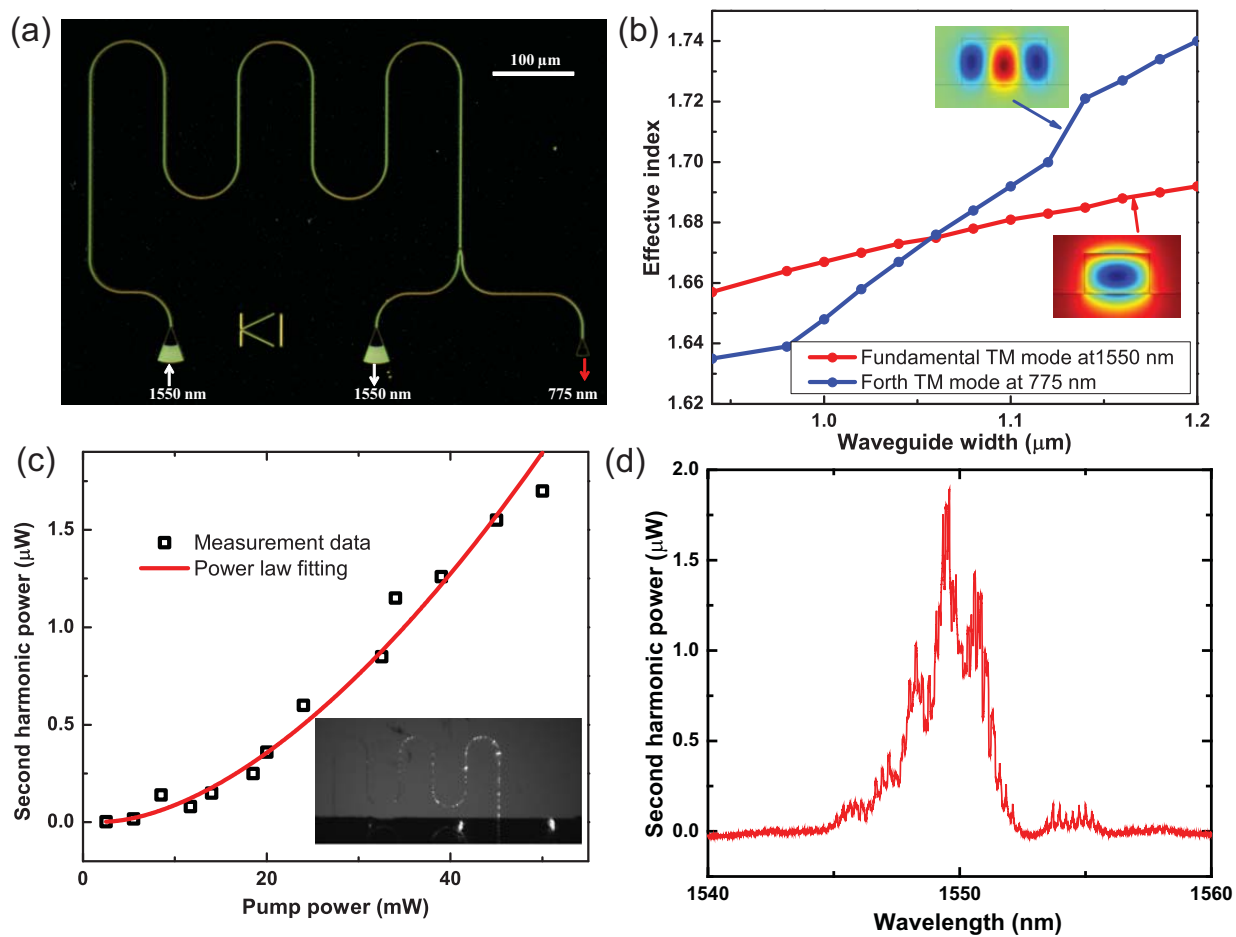

**Figure 6**. (a) An optical microscope image of the 1.6 mm long AlN waveguide; (b) Phase matching is achieved by varying the waveguide width to match the effective index of the fundamental TM mode at the pump wavelength to that of the fourth TM mode of the SH. The optimal waveguide width is found to be 1060 nm. The inset shows the numerically calculated cross-sectional mode profiles (plot in electrical field, *y* component) of the fundamental TM mode at pump wavelength and the fourth TM mode at the SH wavelength. (c) Measured SH power as a function of the pump input power. The solid curve is a power fit to the data. The best fit line gives an exponent of 1.70. Inset: a nanophotonic waveguide glowing at the SH wavelength, captured by a silicon CCD camera (exposure time is 300 ms). The waveguide width is 1.06 μm. The two bright glowing spots near the bottom of the image correspond to the positions of the output couplers at the pump wavelength and the SH wavelength, respectively, since both couplers couple a portion of the SH. (d) The SHG power as a function of the pump wavelength around the phase matching point.

To initiate the nonlinear process, amplified pump light from a tunable CW laser source around 1550 nm wavelength is coupled into the 1.6 mm meander AlN waveguide. When the input wavelength is tuned into the phase matching bandwidth of the waveguide we observe strong SH light at the 775 nm output port. We observe estimated power of 1.7 μW at 775 nm in the waveguide (or equivalently 17 nW coupled out of the chip). The value is obtained by taking into account the loss due to the output grating coupler (-17 dB) and the power splitter, at an average input optical pump power of 50 mW on the waveguide. The propagating power in the feeding waveguide is corrected by the coupling loss at the input grating coupler (-10 dB). From the measured results we obtain a conversion efficiency of -46 dB, which is comparable to previously reported results in gallium nitride ring resonators [12]. While the efficiency is lower than reported values for periodically poled lithium niobate waveguides [41, 42], the device demonstrated here does not rely on poling procedures and is further compatible with ongoing efforts in on-chip CMOS compatible photonics. When we measure the intensity of the SHG light in dependence of pump power on the waveguide as shown in figure 6(c) a near quadratic dependence can be observed. The best fit with a power law reveals an exponent of 1.70. The slight deviation from the exact second order law can be attributed to the SH light propagation loss. The phase matched SHG is strong enough to become visible under an optical microscope due to the scattering from the surface roughness on the waveguides as shown in the inset of figure 6(c). Additionally, we measure the dependence of the SH power on the pump wavelength as shown in figure 6(d), from which we can extract the full width half maximum (FWHM) for the conversion efficiency to be around 4 nm in wavelength. As the figure suggests, when the pump wavelength is close to the phase matching wavelength, the SHG power increases significantly, with a maximum around 1550 nm.

*4.2 Electro-optic tuning in AlN resonators*

AlN photonic circuits also offer electro-optic effects which can be used to enhance passive optical devices with active electrical control. Single crystal AlN is a uniaxial material with 6mm symmetry and *c*-axis as the optic axis. Therefore the electro-optic coefficient matrix of AlN only provides non-vanishing elements of $r_{13}$, $r_{33}$ and $r_{51}$. Because our polycrystalline AlN films have $c$-axis out-of-plane orientation the thin films maintain in-plane isotropy. The electro-optic (EO) coefficient of AlN ($r_{33}$, $r_{13}$~1 pm/V) [43], is comparable with that of GaAs ($r_{14}$=1.5 pm/V) [44] which has been widely employed in commercial phase and amplitude modulators [45, 46]. Compared with LiNbO$_3$ which has a higher electro-optic coefficient ($r_{33}$=33 pm/V), low cost, the potential for chip-scale integration and CMOS compatibility of AlN compensate for its lower EO coefficient.

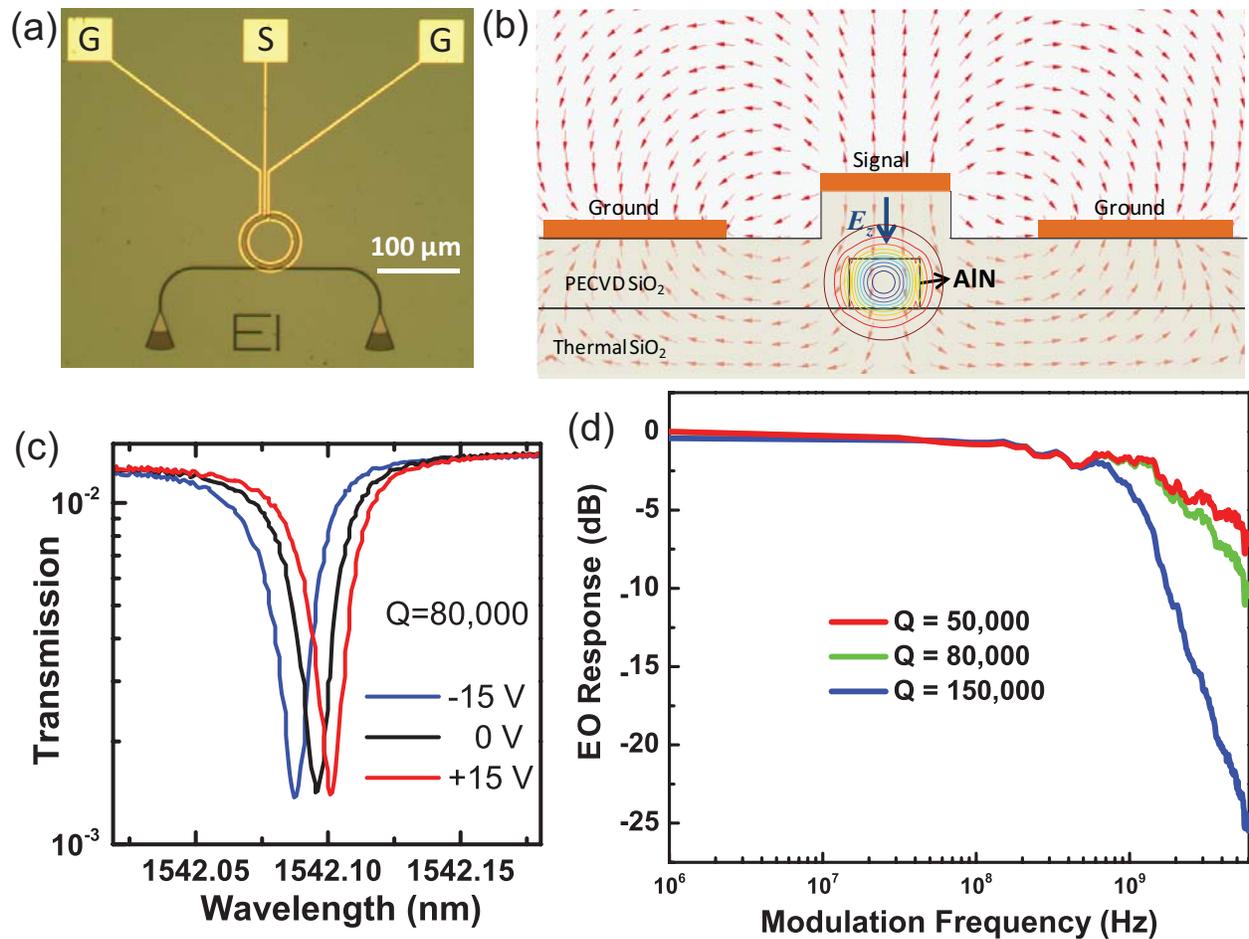

**Figure 7**. (a) A microscopic image of a microring modulator. The inset shows a drawing of the cross section of the microring. The AlN waveguide is capped with 800 nm PECVD silicon dioxide. A set of GSG probe is patterned on the oxide with the centre electrode directly above the AlN waveguide. (b) The simulated RF electric field distribution (arrows) and the optical field (contour plots) distribution. An effective out-of-plane electric field ($E_z$) is generated. (c) A ring resonance near 1542.10 nm with an extinction ratio of 10 dB and quality factor of 80 000 is tuned 15 pm by applying dc bias from -15 V to +15V. (d) The frequency response of the electro-

optic modulation for rings with different $Q$ factors. The speed of the modulation is limited by the cavity photon lifetime.

In order to access AlN's largest EO coefficient [43] $r_{13}$ and $r_{33}$, an out-of-plane electric field ($E_z$) is required, which is normally harder to obtain with planar electrodes. To introduce an effective out-of-plane electric field ($E_z$) overlapping with the guided optical mode, we fabricated a set of ground-signal-ground titanium/gold electrodes (10 nm/200 nm) atop a deposited oxide layer. Figure 7(a) shows an optical micrograph of a fabricated AlN microring modulator circuit with optical input/output ports in the bottom half and RF electrodes in the top half. By applying an electrical field across the electrodes and effective $E_z$ component can be achieved as shown in the simulated field profile in figure 7(b) (vector markers), which overlaps with the optical mode profile (colored contour plot). When a DC bias voltage is applied on the electrodes, the resonance conditions for the microring resonator will change in response to the modified refractive index due to the electro-optic effect. As shown in figure 7(c), we measure a resonance shift of 15 pm when the bias voltage is varied from -15 V to +15 V for a TE-mode resonance around 1542.10 nm with extinction ratio of 10 dB.

To characterize the microring modulator's capability for carrying out signal modulation at high frequency, the input light's wavelength is tuned into one of the ring resonator's resonances. For maximum modulation transduction, the wavelength is biased at half of the transmission point near the resonance, corresponding to the location of the largest slope. Using a network analyzer, we are able to measure the electro-optic modulation in the frequency domain for resonators with quality factors of 50 000, 80 000 and 150 000 respectively. Figure 7(d) shows the EO modulation amplitude ($S_{21}$), revealing a 3 dB electrical bandwidth of 2.3 GHz for the ring resonator with $Q$ of 80 000. For the device with a $Q$ of 50 000 we obtain a cutoff frequency of 3.5 GHz, while the devices with the highest $Q$ of 150 000 yields a cutoff frequency of 1.3 GHz. In these devices, the RC time constant associated with the electrode is around 1 ps, which is less likely to be the main limiting factor of the cutoff frequency. The cavity photon lifetime, on the other hand, calculated by $\tau_{ph} = \lambda Q / 2\pi c$, ($\lambda$ is the wavelength, $Q$ is the quality factor, $c$ is the light speed in vacuum) is around 66 $ps$ for a $Q$ factor of 80 000. As such, the cavity photon lifetime limited cut-off frequency $f_{3-dB} = 1/2\pi\tau_{ph} = 2.4$ GHz is the main limiting factor of our modulator's frequency response, as supported by our network analyzer measurement. The results in figure 7(d) suggest that smaller quality factor can enable faster modulation.

Equivalent experiments were also carried out at visible wavelengths using microring resonators with a reduced diameter of 80 μm. Because of the lower quality factor ($Q$=10 000) in this wavelength regime, the measured 3 dB cutoff frequency is limited by the visible photodiode bandwidth (12.5 GHz) prior to the rolling-off due to the cavity photon lifetime. These results also indicate that higher cutoff frequency can be reached when lower $Q$-resonators are employed.

## 5. Integrated AlN optomechanical resonators

Finally, we design and fabricate monolithically integrated radial contour-mode AlN mechanical ring resonators which simultaneously serve as an optical cavity for sensitive displacement readout. We measure high optical quality factor (loaded $Q$> 125 000) in our suspended AlN ring resonators. The strong optomechanical interaction allows us to resolve the thermomechanical motion of the ring's contour modes with resonance frequencies in excess of GHz at room temperature and ambient pressure. Figure 8(a) shows an SEM micrograph of a suspended AlN ring resonator adjacent to a straight coupling waveguide. As shown in the picture, the ring is supported by four 0.5-μm-wide spokes which are anchored to a central pedestal sitting on top of the underlying oxide. During the release process, a portion of the coupling waveguide is also undercut due to the isotropic nature of the sacrificial oxide etching.

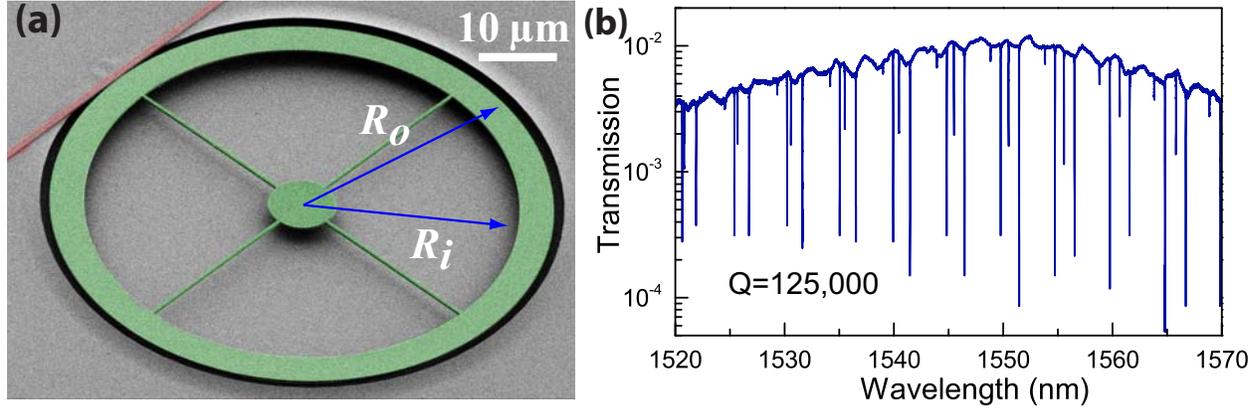

**Figure 8**. (a) False-color scanning electron micrograph of the fabricated AlN ring resonator (in green). The coupling waveguide is denoted in red. (b) Optical transmission spectrum of a released AlN optomechanical ring resonator measured in air with -20 dBm of input power before the input grating coupler. The free spectral range is 5.0 nm. Critical coupled quality factor is near 125 000.

To realize GHz mechanical resonances, we focus on the radial-contour modes of ring resonators with a geometry similar to electrostatically actuated "hollow-disk" resonators [47]. These radial-contour modes expand/contract the ring width and their resonance frequencies can be approximated by $f = \frac{m}{2W}\sqrt{\frac{E}{\rho}}$, $m = 1, 2, 3, \cdots$, where $W$ is the width of the ring, $E$ and $\rho$ are the Young's modulus ($E_{AlN} = 330$ GPa) and the density ($\rho_{AlN} = 3330$ kg/m$^3$) of the material, respectively. To design a target 1 GHz resonance frequency with $m = 1$, the ring width needs to be $W = 5$ μm. Furthermore, in order to minimize the anchor loss to the substrate and maximize the mechanical $Q$, the spoke length $L_{sp}$, defined as the distance from the wheel centre to the ring attaching point, should be an odd multiple of a quarter wavelength of the spoke's longitudinal mode: $L_{sp} = (2n-1)\frac{\lambda_a}{4} = \frac{2n-1}{4f}\sqrt{\frac{E}{\rho}}$, $n = 1, 2, 3 \ldots$ In this study, we choose a spoke length $L_{sp} =$ $13\lambda_a/4 = 32.6\ \mu m$, and the inner and outer radii are $R_i = L_{sp} = 32.6\ \mu m$ and $R_o = 37.6\ \mu m$ accordingly.

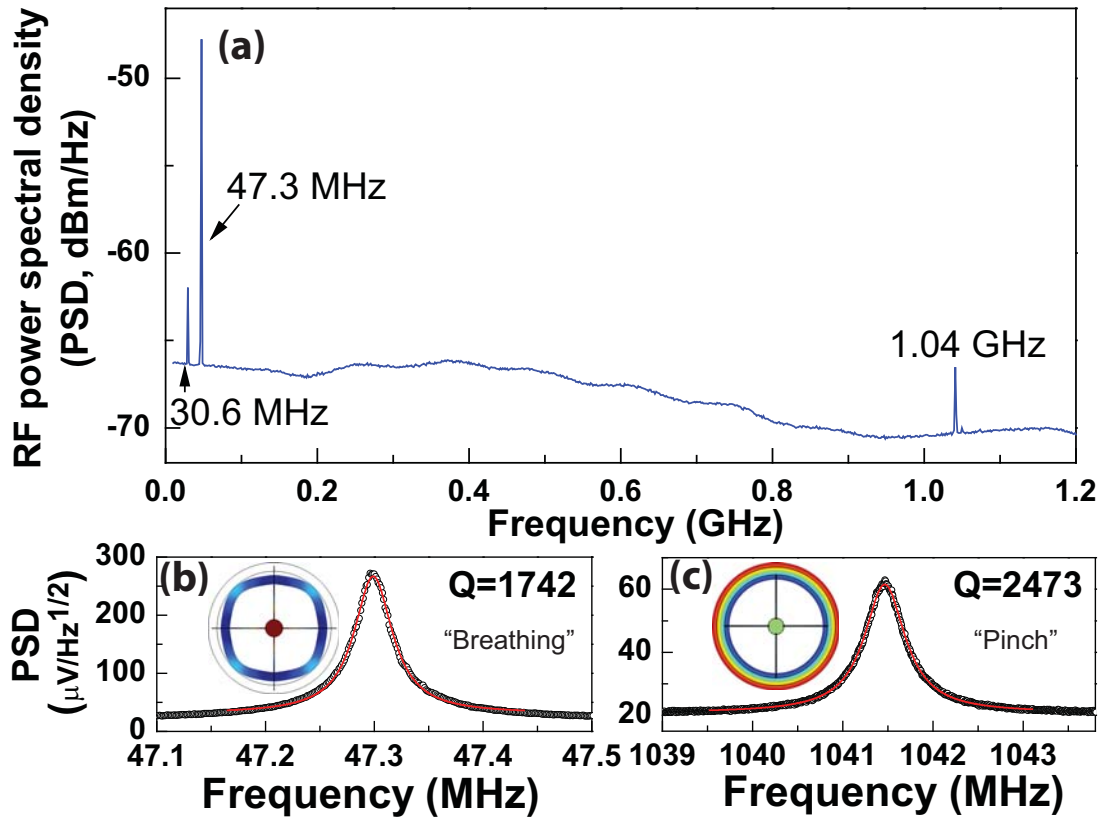

**Figure 9**. (a) Optically transduced RF spectrum of the mechanical modes of the AlN ring resonator measured in air with an input power of 3 dBm. Three mechanical modes are found at 30.6 MHz, 47.3 MHz, and 1.04 GHz. (b) and (c) are the zoom-in spectra of the resonance at 47.3 MHz ("breathing" mode), and 1041.5 MHz ("pinch" mode), respectively. The fitted (red lines) quality factors are 1742 and 2473 measured in air respectively. The insets in (b) and (c) are the numerically simulated displacement profile for the respective mode.

Figure 8 (b) shows the transmission spectrum of a fabricated released ring resonator measured at low input power (-20 dBm before the input grating coupler). Because the width of the ring supports multiple modes, several sets of resonances show up with different extinction ratios. In order to optically transduce the mechanical modes, we tune the input laser to the maximum slope of an optical resonance and record the RF power spectrum of the optical transmission. To resolve the intrinsic mechanical $Q$ factors, we keep the laser input power low enough (+3 dBm) that the optomechanical back-action is negligible. Under this condition, the light transmitted past the cavity is amplified by an EDFA (Pritel FA-20) and then detected by a high-speed InGaAs photoreceiver with 1 GHz bandwidth (New Focus 1611). The output of the photoreceiver is analyzed by a spectrum analyzer. Figure 9(a) shows the RF spectrum of the mechanical modes of the wheel resonator measured in air. The 1st contour mode ("breathing" mode) is detected at 47.3 MHz with a measured $Q$ of 1742 [figure 9(b)]. The 2nd contour mode ("pinch" mode) at 1.04 GHz has a slightly higher measured $Q$ of 2473 [figure 9(c)].In addition, a 30.64-MHz resonance is observed and identified as one of the wineglass modes (circumferential number n = 4, "square" mode [48]) with a measured $Q$ of 1200.

By comparing the expected displacement noise with the measured RF spectral density, we can calibrate the amplitude of the Brownian motion for each mechanical mode. The displacement sensitivity corresponds to the noise floor of the spectrum. (The spectral density of the thermomechanical displacement noise at resonance frequency is $S_x^{1/2} = \sqrt{4k_\mathrm{B}TQ/m_\mathrm{eff}(2\pi f)^3}$, where $k_\mathrm{B}$ is the Boltzmann constant, $T$ is the absolute temperature (300 K), $Q$ is the mechanical quality factor, $m_\mathrm{eff}$ is the effective modal mass, and $f$ is the resonance frequency. The effective modal mass is defined as $m_\mathrm{eff} = \dfrac{m_0}{u_\mathrm{max}^2 V} \iiint u(x,y,z)^2 \mathrm{d}x\mathrm{d}y\mathrm{d}z$, where $m_0$ is the physical mass of the ring, $u(x,y,z)$ is the displacement, $u_{max}$ is the maximum displacement, $V$ is the volume of the ring.) Table 1 summarizes the simulated and experimental resonance frequencies, together with the simulated modal effective mass $m_{eff}$ and the calibrated displacement sensitivity $S_x^{1/2}$ of the three measured modes. The table shows that the experimental frequencies are in good agreement with their simulated values. One of the strategies to further enhance the optomechanical coupling is to employ rings with smaller radii, as the coupling strength $g_{om} = \mathrm{d}\omega/\mathrm{d}R$ (currently $g_{om} = 5.3$ GHz/nm) is inversely proportional to the radius $R$.

**Table 1.** Properties of the observed mechanical modes: simulated (superscript $s$), and experimental (superscript $e$) frequencies, effective modal mass ($m_{eff}$), measured quality factor ($Q_m$) and displacement sensitivity ($S_x^{1/2}$) of the ring resonator with inner radius $R_i$ = 32.6 μm and outer radius $R_o$ = 37.6 μm.

| Mode type | $f^{(s)}$ (MHz) | $f^{(e)}$ (MHz) | $m_{\mathrm{eff}}^{(s)}$ (ng) | $Q_{\mathrm{m}}^{(e)}$ (air) | $S_x^{1/2}$ (m/√Hz) |
|---|---|---|---|---|---|
| "Square" | 30 | 30.6 | 0.63 | 1200 | $3.5\times10^{-16}$ |
| "Breathing" | 47 | 47.3 | 1.14 | 1742 | $9.3\times10^{-17}$ |
| "Pinch" | 1050 | 1041.5 | 0.42 | 2473 | $6.2\times10^{-18}$ |

## 6. Conclusions

In summary, we have explored AlN thin film as a new material system for building optomechanical and nonlinear optical devices that assist on-chip wavelength conversion and electro-optic modulation. High optical $Q$ (125 000) and mechanical $Q$ (2473) are observed in AlN optomechanical ring resonators vibrating at GHz frequencies. Compared with conventional materials for integrated cavity optomechanical systems (such as silicon and silicon nitride), AlN offers many unique optical and mechanical properties. Its wide bandgap (6.2 eV) provides superior suppression of two-photon absorption and thus enables more stable optical resonators with high power handling capability. More importantly, AlN has a large piezoelectric coefficient and an intrinsic electro-optic Pockels' effect [43], which make AlN resonators an interesting platform to implement tunable, electrically driven and optically read out oscillator systems.

The potential of AlN photonic technologies lies in the fact that wafer-scale deposition of high quality films can be conducted in a mature procedure compatible with CMOS manufacturing. The nature of the AlN sputtering process also makes it possible to design three dimensional and multi-layer structures which can enable more flexible and efficient integration. Using our fully CMOS-compatible AlN photonic circuits, we have thus implemented a promising platform for integrated optomechanical circuits with on chip signal processing capability.

## Acknowledgments

The authors acknowledge the funding support from DARPA ORCHID program through a grant from the Air Force Office of Scientific Research (AFOSR C11L10831), managed by Dr. J. R. Abo-Shaeer, the Packard Foundation, and the NSF grant through MRSEC DMR 1119826, and a NSF CAREER award. W.H.P.P. acknowledges support by the DFG Grant PE 1832/1-1.